\documentclass[preprint,12pt]{elsarticle}
\usepackage{epsfig}
\usepackage{amssymb}
\usepackage{amsmath}

\journal{Solid State Sciences}

\begin{document}

\begin{frontmatter}

\title{Optical and transport properties of Ba(Fe$_{1-x}$Ni$_x$)$_2$As$_2$ films}

\author[label1]{Yurii A. Aleshchenko \corref{cor1}}
\ead{aleshchenkoya@lebedev.ru}
\cortext[cor1]{Corresponding author}

\affiliation[label1]{organization={P.N. Lebedev Physical Institute, Russian Academy of Sciences},
            addressline={Leninskiy Prospekt 53},
            city={Moscow},
            postcode={119991},
            state={},
            country={Russia}}

\author[label1]{Andrey V. Muratov}

\author[label2]{Elena S. Zhukova}

\author[label2]{Lenar S. Kadyrov}

\author[label2]{Boris P. Gorshunov}

\affiliation[label2]{organization={Moscow Institute of Physics and Technology},
            addressline={Institutskiy per. 9},
            city={Dolgoprudny},
            postcode={141700},
            state={Moscow Region},
            country={Russia}}

\author[label3,label4]{Giovanni A. Ummarino}

\affiliation[label3]{organization={Istituto di Ingegneria e Fisica dei Materiali, Dipartimento di Scienza Applicata e Tecnologia, Politecnico di Torino},
            addressline={Corso Duca degli Abruzzi 24},
            city={Torino},
            postcode={10129},
            state={},
            country={Italy}}

\affiliation[label4]{organization={National Research Nuclear University MEPhI (Moscow Engineering Physics Institute)},
            addressline={Kashirskoe shosse 31},
            city={Moscow},
            postcode={115409},
            state={},
            country={Russia}}

\author[label1]{Ilya A. Shipulin}

\begin{abstract}
The broad-band optical spectroscopy was used to study the optical and the hidden transport properties of the Ba(Fe$_{1-x}$Ni$_x$)$_2$As$_2$ superconducting films with different Ni contents. The normal state data were analyzed using a Drude-Lorentz model with two Drude components: narrow and broad ones. In the superconducting state, two gaps with $2\Delta _{0}^{(1)}/k_{B}T_{c}=1.57$ and $2\Delta _{0}^{(2)}/k_{B}T_{c}=3.48$ are formed for the Ba(Fe$_{0.965}$Ni$_{0.035}$)$_2$As$_2$ films, while for the Ba(Fe$_{0.95}$Ni$_{0.05}$)$_2$As$_2$ films these characteristic ratios are 1.88--2.08 and 3.66--4.13. Both gaps are formed from the narrow Drude component, whereas the broad Drude component remains ungapped. The calculated from infrared data total dc resistivity of the films with Ni contents $x=0.05$ and $x=0.08$ as well as the low-temperature scattering rate for the narrow Drude component show a hidden Fermi-liquid behavior. On the contrary, the films with $x=0.035$ manifest a hidden non-Fermi-liquid behavior. The Allen theory generalized to a multiband systems was applied to the analysis of the temperature dependences of a resistivity of the Ba(Fe$_{1-x}$Ni$_x$)$_2$As$_2$ films. The change of total electron-boson coupling and representative energy in the normal state versus the superconducting state for this system was shown to be typical of other iron-based superconducting materials as well as high-temperature superconducting cuprates.
\end{abstract}

%%Graphical abstract
%\begin{graphicalabstract}
%\includegraphics{grabs}
%\end{graphicalabstract}

\begin{highlights}
\item The broad-band optical spectra of the Ba(Fe$_{1-x}$Ni$_x$)$_2$As$_2$ superconducting films with different Ni contents were obtained for the first time.
\item The values of the superconducting gaps were obtained for the films with Ni contents $x=0.035$ and $x=0.05$.
\item The temperature dependences of the plasma frequencies, optical conductivities, static scattering rates, and resistivities in the normal state deduced from the optical spectra of the  Ba(Fe$_{1-x}$Ni$_x$)$_2$As$_2$ films have been obtained.
\item The Fermi-liquid behavior was established for the films with $x=0.05$ and $x=0.08$, meanwhile the films with $x=0.035$ manifested the non-Fermi-liquid behavior.
\item The Allen theory generalized to a multiband systems has been applied to the analysis of the temperature dependences of a resistivity of the Ba(Fe$_{1-x}$Ni$_x$)$_2$As$_2$ films and the change of total electron-boson coupling, as well as representative energy in the normal state versus the superconducting state for this system was shown to be typical of other iron-based superconducting materials as well as high-temperature superconducting cuprates.
{\sloppy

    }
\end{highlights}

%% Keywords
\begin{keyword}
Multiband superconductivity, Fe-based superconductors, Allen model, THz spectroscopy, IR spectroscopy, ellipsometry
\PACS 74.70.Xa, 74.20.Fg, 74.25.Kc, 74.20.Mn, 74.25.Gz

\end{keyword}

\end{frontmatter}

\section{Introduction}

Iron-based superconductors (IBS) have been intensively studied since their discovery~\cite{Kamihara1}. They belong to multiband systems due to the presence of multiple orbitals at the Fermi level~\cite{Lebegue,Ding1,Kondo,Mazin,Subedi}. Such systems may exhibit intriguing  features such as multiple superconducting gaps~\cite{Ding1,Kondo,Wu,Heumen,Dressel1,Dai} and non-trivial gap symmetries~\cite{Bang,Grinenko}. Among these materials, electron and hole doped BaFe$_2$As$_2$ (Ba122) compounds deserve special attention for a number of reasons: first, the relatively simple growth technology at ambient pressure; second, the possibility of obtaining really large and high quality single crystals; and finally, the possibility of inducing superconductivity in different ways (i.e., by applying external pressure, substitution of each atomic site, or a combination of both pathways), resulting in comparable phase diagrams~\cite{Canfield}. In general, Ba122 possesses a layered crystal structure in which Fe--As layers are separated by Ba-based spacers along the crystallographic $c$-direction. At room temperature, Ba122 is a paramagnetic metal with a tetragonal structure. The resistivity in the planes decreases with temperature down to an  anomalous drop when the material undergoes a magnetic transition at $T_N\approx 138$~K to a spin-density-wave (SDW)-like antiferromagnetic (AFM) ground state that is also accompanied by a structural transition to an orthorhombic phase~\cite{Rotter1}. However, when Fe is partially replaced by Ni, AFM order is suppressed, and a superconducting (SC) state emerges~\cite{Wang1}. Such behavior is very clearly observed for both Ba(Fe$_{1-x}$Ni$_x$)$_2$As$_2$ single crystals~\cite{Lee} and films~\cite{Yoon,Shipulin2018}, the latter, incidentally, have a high-quality epitaxial structure as well as electronic and SC properties comparable, and in some cases even superior, to single crystals. However, the optical studies of Ba(Fe$_{1-x}$Ni$_x$)$_2$As$_2$ films have been few and limited only by the films with the optimal Ni content ($x\simeq 0.05$)~\cite{Yoon,Yuriisup,Aleshchenko2021}, and only in our study~\cite{Aleshchenko2021} have the SC gap values for $x=0.045$ been determined experimentally. Also, the issue of SC gaps for the optimally Ni-doped Ba122 single crystals was only briefly addressed in~\cite{Dressel1,Wu1}, despite the fact that the optical spectroscopy is a powerful technique to investigate the band structure and charge dynamics of the material. Moreover, the broad-band optical studies of these materials are lacking.

In this paper, for the first time we undertake the broad-band optical studies of Ba(Fe$_{1-x}$Ni$_x$)$_2$As$_2$ films with Ni contents $x=0.035$ (underdoped), $x=0.05$ (optimally doped), and $x=0.08$ (overdoped) combining the following techniques: the terahertz (THz) time-domain spectroscopy, the Fourier-transform infrared (FT-IR) spectroscopy, and the spectroscopic ellipsometry. In our optical studies, we have revealed the hidden transport properties of the films and compare them with the results of dc resistivity measurements analyzed within the Allen model~\cite{Allen}.

\section{Experiment}
The high-quality epitaxial films of Ba(Fe$_{1-x}$Ni$_x$)$_2$As$_2$ ($x=0.035$, 0.05, and 0.08) with the thicknesses of 100--150~nm were prepared  by the standard pulsed laser deposition (PLD) method on double-polished (001) CaF$_2$ substrates. The base pressure in the chamber was better than $1\times 10^{-8}$~mbar and increased slightly to $2\times 10^{-7}$ mbar during deposition. More details on the sample preparation can be found in~\cite{Shipulin2018,Richter1,Richter2,Shipulin2}. All as-produced films have a fairly smooth surface morphology without droplets and a crystalline structure with $c$-axis orientation. The composition of the films was checked by EDX measurements.The resistivity measurements of the films were performed with the standard four-probe technique in the Van der Pauw scheme.

The THz spectroscopic measurements were performed in the transmission mode with the Menlo Systems Tera K15 pulsed time-domain THz spectrometer within the range of 10--55 cm$^{-1}$ (wavelengths 1 mm -- 200 $\mu $m) over a wide temperature range down to $T=4$ K with a home-made optical cryostat. The measured transmission coefficient and the phase shift of the electromagnetic radiation passed through the film on a substrate provide direct information on the spectra of the real and imaginary parts of the complex dielectric permittivity $\tilde\varepsilon =\varepsilon _{1}(\omega)+i\varepsilon _{2}(\omega )$ and the complex optical conductivity $\tilde\sigma (\omega )=\sigma _{1}(\omega )+i\sigma _{2}(\omega )$ without the need of Kramers-Kronig transformation~\cite{Pracht}.

The IR reflectivity spectra were measured in the near-normal incidence geometry in the spectral range of 25--10~000~cm$^{-1}$ (400--1~$\mu $m). The variable-angle spectroscopic ellipsometry was carried out in the near-infrared to ultraviolet range from 4000 to 50 000 cm$^{-1}$ (2.5~$\mu$m -- 200 nm). For the analysis of the experimental data a two-layer system Ba(Fe$_{1-x}$Ni$_x$)$_2$As$_2$  film -- CaF$_2$ substrate of finite thickness was considered. A detailed extended account on both these experimental techniques and data analysis may be found in our previous paper~\cite{Aleshchenko2021} and in the Supplemental Materials to it.

\section{Experimental results}

The temperature dependences of dc resistivity of the Ba(Fe$_{1-x}$Ni$_x$)$_2$As$_2$ films are shown in Fig.~1. It is worth noting the rather narrow (about 2~K) transition to the SC state for all the obtained films, as well as resistivities comparable to single crystals~\cite{Wang}, which also indicates the high structural quality of the films. The SC transition temperatures ($T_c$) evaluated at 90$\%$ of the normal state resistance are 21.1~K, 21.6 K, and 10.3 K for $x=0.035$, 0.05, and 0.08, respectively. In the case when the critical temperature is defined as the temperature of maximum of the derivative of resistivity with respect to temperature, the $T_c$ values are the following: 20.27~K ($x = 0.035$), 20.36~K ($x = 0.05$), and 9.27~K ($x = 0.08$).

\begin{figure}[!ht]
\includegraphics[width=8.3cm]{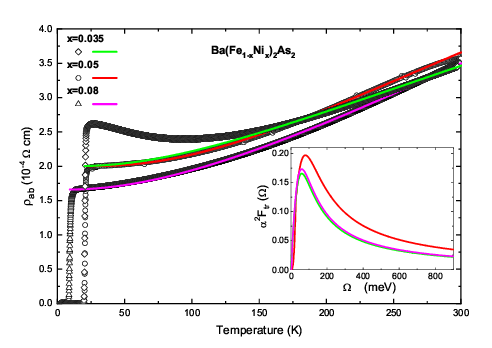}
\centering
\caption{(Color online) Temperature dependence of the resistivity for the Ba(Fe$_{1-x}$Ni$_x$)$_2$As$_2$ films (black open symbols). The solid curves represent the fit obtained using the Allen theory (see sections 4 and 5 for more details). The inset shows the spectral functions of the antiferromagnetic spin fluctuations in the normal state normalized to unity (see sections 4 and 5).}
%\label{Fig_1}
\end{figure}

Figures 2--4 display the broad-band reflectivity spectra $R(\nu )$ of three Ba(Fe$_{1-x}$Ni$_x$)$_2$As$_2$ films  with $x=0.035$, 0.05, and 0.08 at some representative temperatures (panels (a)) together with the real parts of the optical conductivity derived from the Drude--Lorentz analysis (panels (b)). The fit to the reflectivity curves based on the Drude--Lorentz analysis is also shown and will be discussed below. In all spectra, the reflectivity $R(\nu )$ exhibits a metallic response in the normal state in both frequency and temperature, i.e., $R(\nu )$ is enhanced at low frequencies below $\sim 480$~cm$^{-1}$ ($x=0.035$), $\sim 545$~cm$^{-1}$ ($x=0.05$), and $\sim 660$~cm$^{-1}$ ($x=0.08$) with decreasing temperature. A sharp increase in the reflectivity below 52--57~cm$^{-1}$ (6.4--7.1~meV), which is associated with the formation of the SC gap due to electron pairing, is observed below $T_c$ for the films with $x=0.035$ and $x=0.05$. For the Ba(Fe$_{0.92}$Ni$_{0.08}$)$_2$As$_2$ films, the increase in $R(\nu )$ below $T_c$ is not pronounced due to the smaller SC gap for this Ni content and the limited frequency  range of our measurements. In the temperature range of 4--100~K, the reflectivity curves almost overlap with each other above 100~cm$^{-1}$. The decrease in $R(\nu )$ becomes steep above 300~cm$^{-1}$. Above 5700~cm$^{-1}$ the difference between all spectra, including those taken at 100--300~K, becomes almost imperceptible. The acquired spectra also contain two sharp features at $\sim 94$~cm$^{-1}$ and $\sim 260$~cm$^{-1}$, which correspond to the contributions of phonons of the CaF$_2$ substrate and symmetry-allowed IR-active $E_u$ phonons in the $ab$-plane of the films~\cite{Akrap}.

\begin{figure}[!ht]
\includegraphics[width=8.3cm]{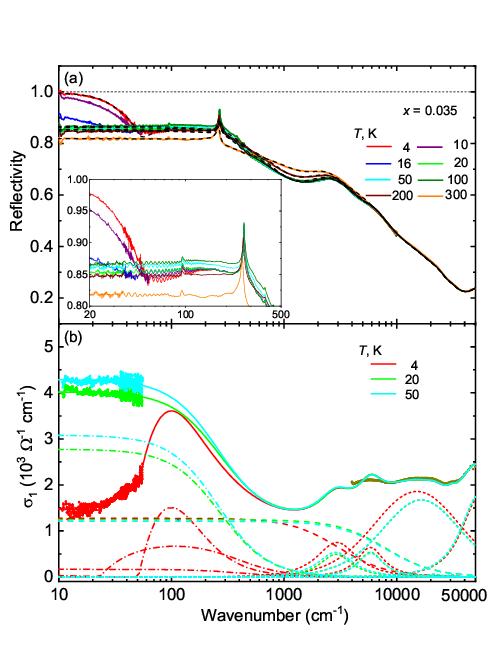}
\centering
\caption{(Color online) (a) The reflectivity spectra of the Ba(Fe$_{0.965}$Ni$_{0.035}$)$_2$As$_2$ film at various temperatures together with the Drude--Lorentz fit (dashed lines). The inset shows an expanded plot of reflectivity spectra below 500~cm$^{-1}$. (b) Optical conductivity data together with three representative Drude-Lorentz fits for the  Ba(Fe$_{0.965}$Ni$_{0.035}$)$_2$As$_2$ film at temperatures of 4, 20, and 50 K, respectively. Conductivity data below 55~cm$^{-1}$ (dots) were obtained with the THz time-domain measurements. Corresponding reflectivity was calculated using THz conductivity and permittivity. Reflectivity (conductivity) data above 4000~cm$^{-1}$ were calculated directly from ellipsometry measurements. Dash-dotted lines: $D_1$ contributions. Dashed lines: $D_2$ contributions. Short-dashed lines: $L_1$--$L_6$ Lorentzian contributions. The solid lines represent the total optical conductivity spectra.}
%\label{Fig_2}
\end{figure}

We applied a two-layer modeling of the film/substrate system to the analysis of the experimental reflectivity together with the directly measured dielectric permittivity and optical conductivity from the THz and ellipsometry data of the Ba(Fe$_{1-x}$Ni$_x$)$_2$As$_2$ films. The optical response of the films was modeled within a two-Drude approach proposed earlier in~\cite{Wu1} to study properties of multiband IBS. In this case, the optical conductivity of the films in the normal state is modeled by the two Drude components, narrow (D$_1$) and broad (D$_2$) as well as a set of Lorentz components representing the interband transitions. According to the Drude-Lorentz model, the complex dielectric function $\tilde\varepsilon (\omega)=\varepsilon _1(\omega)+i\varepsilon _2(\omega)$ can be represented as
$$
\tilde\varepsilon (\omega)=\varepsilon _{\infty }-\sum _{i=1,2}\frac{\omega _{Di,p}^2}{\omega (\omega +i\gamma _{Di})}+\sum _j\frac{\omega _{j,p}^2}{\omega _j^2-\omega ^2-i\gamma _j\omega},
$$
where $\varepsilon _\infty $ is the background dielectric function, which comes from contribution of the high frequency absorption bands, the subscript D$_1$ (D$_2$) denotes the narrow (broad) Drude component, $\omega _{Di,p}$ is the Drude plasma frequency, $\gamma _{Di}$ is the (average) elastic scattering rate among free charge carriers, $\omega _{j,p}$, $\omega _j$, and $\gamma _j$ are the plasma frequency, the center frequency, and the damping of the $j^{th}$ Lorentz component, respectively. The optical conductivity can be related to the dielectric function as $\tilde\sigma (\omega )=\sigma _1(\omega)+i\sigma _2(\omega )=i\omega [\varepsilon _\infty -\tilde\varepsilon (\omega )]/4\pi $.

\begin{figure}[!ht]
\includegraphics[width=8.3cm]{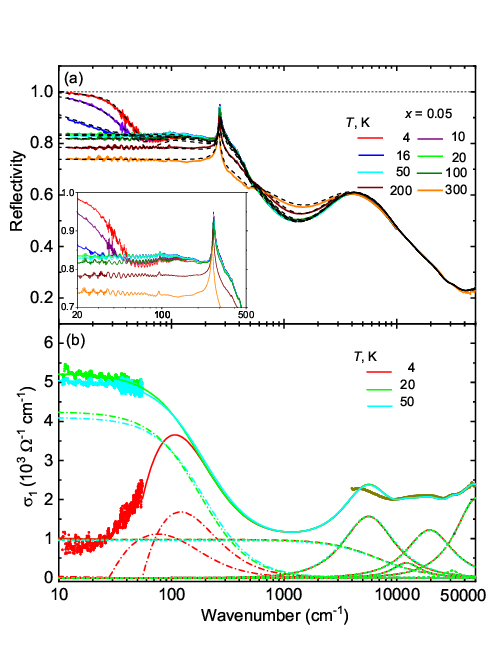}
\centering
\caption{(Color online) (a) The reflectivity spectra of the Ba(Fe$_{0.95}$Ni$_{0.05}$)$_2$As$_2$ film at various temperatures together with the Drude--Lorentz fit (dashed lines). The inset shows an expanded plot of reflectivity spectra below 500~cm$^{-1}$. (b) Optical conductivity data together with three representative Drude-Lorentz fits for the Ba(Fe$_{0.95}$Ni$_{0.05}$)$_2$As$_2$ film at temperatures of 4, 20, and 50 K, respectively. Conductivity data below 55~cm$^{-1}$ (dots) were obtained with the THz time-domain measurements. Corresponding reflectivity was calculated using THz conductivity and permittivity. Reflectivity (conductivity) data above 4000~cm$^{-1}$ were calculated directly from ellipsometry measurements. Dash-dotted lines: $D_1$ contributions. Dashed lines: $D_2$ contributions. Short-dashed lines: $L_1$--$L_6$ Lorentzian contributions. The solid lines represent the total optical conductivity spectra.}
%\label{Fig_3}
\end{figure}

\begin{figure}[!ht]
\includegraphics[width=8.3cm]{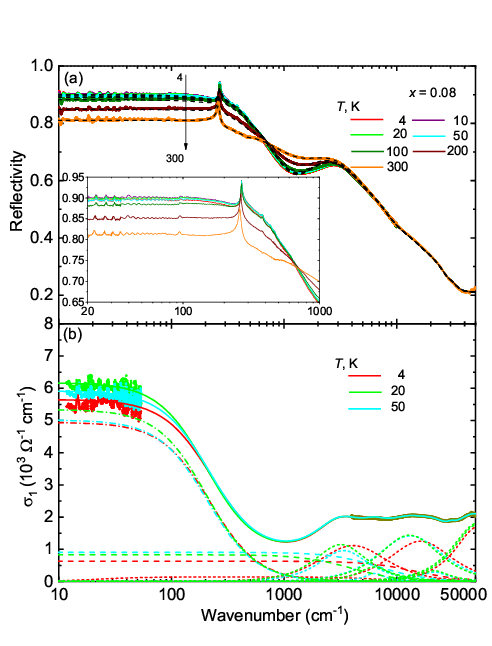}
\centering
\caption{(Color online) (a) The reflectivity spectra of the Ba(Fe$_{0.92}$Ni$_{0.08}$)$_2$As$_2$ film at various temperatures together with the Drude--Lorentz fit (dashed lines). The inset shows an expanded plot of reflectivity spectra below 1000~cm$^{-1}$. (b) Optical conductivity data together with three representative Drude-Lorentz fits for the Ba(Fe$_{0.92}$Ni$_{0.08}$)$_2$As$_2$ film at temperatures of 4, 20, and 50 K, respectively. Conductivity data below 55~cm$^{-1}$ (dots) were obtained with the THz time-domain measurements. Corresponding reflectivity was calculated using THz conductivity and permittivity. Reflectivity (conductivity) data above 4000~cm$^{-1}$ were calculated directly from ellipsometry measurements. Dash-dotted lines: $D_1$ contributions. Dashed lines: $D_2$ contributions. Short-dashed lines: $L_1$--$L_6$ Lorentzian contributions. The solid lines represent the total optical conductivity spectra.}
%\label{Fig_4}
\end{figure}

The Drude term in the Drude-Lorentz model should be replaced by the Zimmermann term~\cite{Zimmermann} below $T_c$. The latter  generalizes the standard BCS Mattis-Bardeen model~\cite{Tinkham1996,Dressel2002} to arbitrary $T$ and $\gamma $ values, with the inclusion of two additional parameters, the SC gap ($\Delta $) and the ratio $T/T_c$.

The results of the modeling for the real part of the optical conductivity spectra at various temperatures based on the Drude-Lorentz analysis are shown in Figs.~2(b), 3(b), and 4(b). Dots in the ranges of 4000--50~000~cm$^{-1}$  and 10--55~cm$^{-1}$ are the optical conductivities calculated directly from the ellipsometry and the THz experiments, respectively. The phonon peaks at $\sim 94$~cm$^{-1}$ and $\sim 260$~cm$^{-1}$ were not included into the film model, which, nevertheless, provided a good quality of the fit. These peaks were included into the substrate model.

In the normal state, the optical conductivity of the films is decomposed into D$_1$ and D$_2$ Drude terms as well as a set of six Lorentzians. The observation of narrow and broad terms in the two-Drude analysis is in agreement with other optical studies, and appears to be a general feature for most of the IBS having multiband structure~\cite{Wu1}. In particular, for the  Ba(Fe$_{0.95}$Ni$_{0.05}$)$_2$As$_2$ films, the D$_1$ component increases in strength and narrows with decreasing temperature, whereas the  D$_2$ component as well as Lorentzian contributions remain virtually unchanged (see Fig.~3(b)). A strong absorption centered around  5000~cm$^{-1}$ can be assigned to the interband transition from As \textit {p} to Fe \textit {d} states~\cite{Yin}. However, for the Ba(Fe$_{1-x}$Ni$_x$)$_2$As$_2$ films with $x=0.035$ and $x=0.08$, the D$_2$ Drude components become anomalously broad with decreasing temperature and overlap with the strong Lorentz contributions in the ranges of 3000--6000~cm$^{-1}$ and 13~500--16~500~cm$^{-1}$, thus preventing an unambiguous determination of the D$_2$ plasma frequency and damping, as can be clearly seen in Figs.~2(b) and 4(b). To overcome this problem, we can act in two different ways. First, with consideration of the Drude-Lorentz analysis for the $x=0.05$ film, we have used the fixed Lorentz contributions to analyse the temperature dependences of the optical conductivity of the films with $x=0.035$ and $x=0.08$. Such an approach is also justified by the fact that the Lorentz contributions have negligible temperature dependences for the Ba(Fe$_{0.975}$Ni$_{0.025}$)$_2$As$_2$ and Ba(Fe$_{0.95}$Ni$_{0.05}$)$_2$As$_2$ single crystals~\cite{Lee},  Ba(Fe$_{0.95}$Ni$_{0.05}$)$_2$As$_2$ films~\cite{Yoon} as well as for the Ba(Fe$_{0.955}$Ni$_{0.045}$)$_2$As$_2$ films studied by us~\cite{Aleshchenko2021}. Second, we have fixed the scattering rate of the broad Drude component $\gamma _{D2}$ at 3000~cm$^{-1}$ in our calculations for the $x=0.035$ and $x=0.05$ films, which is comparable to that found in the aforementioned works. Moreover, the results of the works~\cite{Lee,Yoon} demonstrate negligible temperature dependences of $\gamma _{D2}$. The results of the first approach are presented in Figs.~5 and 7 of this paper, while the temperature dependences of the Drude parameters in the normal state of these films obtained within the second approach are shown in Figs.~8 and 9 of the Appendix. We will show below that both approaches do not affect our main conclusions.

Upon passing through the SC transition, a gap-like structure is formed in the narrow Drude term for the Ba(Fe$_{1-x}$Ni$_x$)$_2$As$_2$  films with $x=0.035$ and $x=0.05$. In the case of Ba(Fe$_{0.92}$Ni$_{0.08}$)$_2$As$_2$ films, no gap is observed because the critical temperature $T_c$ for this composition is too low and accordingly, the SC gap is smaller. Figure~10 in the Appendix illustrates the behavior of the imaginary part of optical conductivity $\sigma _2$ for the films with $x=0.035$ and $x=0.05$. The $1/\omega $ divergence of the $\sigma _2$ inherent to the SC state is evident at the temperature of 4~K. Assuming that two isotropic SC gaps open simultaneously below $T_c$, we obtain a better description of the low-frequency optical conductivity $\sigma _1$ with $2\Delta _{0}^{(1)}/hc=23$~cm$^{-1}$ (2.85~meV) and $2\Delta _{0}^{(2)}/hc=51$~cm$^{-1}$ (6.3~meV) for $x=0.035$ and $2\Delta _{0}^{(1)}/hc=(29.7\pm 1.5)$~cm$^{-1}$ ($3.7\pm 0.2$)~meV and $2\Delta _{0}^{(2)}/hc=(58.5\pm 3.5)$~cm$^{-1}$ ($7.25\pm 0.4$)~meV for $x=0.05$. Here, $h$ is Planck's constant and $c$ is the speed of light.

\begin{figure}[!ht]
\includegraphics[width=8.3cm]{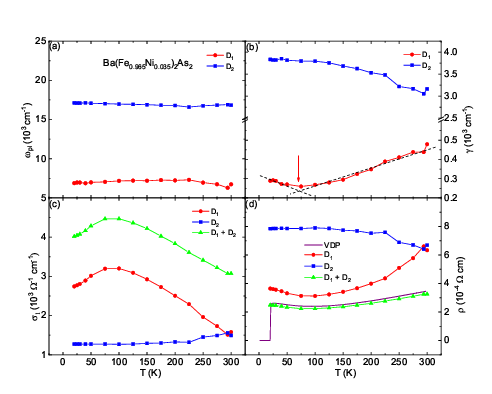}
\centering
\caption{(Color online) The fitting parameters $\omega _{Di,p}$ (a) and $\gamma _{Di}$ (b) of two Drude modes D$_i$ ($i=1,2$) of the Ba(Fe$_{0.965}$Ni$_{0.035}$)$_2$As$_2$ films obtained with the fixed Lorentz contributions. The dashed lines in the panel (b) are linear fits to the scattering rate for the narrow Drude component. The red arrow marks the magnetic transition. Also shown are the calculated dc conductivities $\sigma _{dc,i}(T)$ (c) and dc resistivities $\rho _{i}(T)$ (d) including the total conductivity and resistivity as functions of temperature. The solid purple line depicts the temperature dependence of resistivity obtained from Van der Pauw measurements.}
%\label{Fig_5}
\end{figure}

The temperature dependences of parameters, such as the plasma frequency $\omega _{Di,p}$, the static scattering rate $\gamma _{Di}$, the dc conductivity ($\sigma _{dc}$), and the dc resistivity $\rho $, of two Drude contributions D$_1$ and D$_2$ for the Ni-doped Ba122 films in the normal state are shown in Figs.~5--7. One can see that for the underdoped films ($x=0.035$) the static scattering rate of the D$_1$ component decreases linearly, while cooling down to the magnetic transition temperature $T_N\simeq 70$~K (marked with a red arrow) from 300~K, and then increases linearly below this temperature. The same behavior for the films with $x=0.035$ is also seen in Fig.~8 of the Appendix. According to NMR measurements~\cite{Zhou}, $T_{N}=48$~K for the Ba(Fe$_{0.965}$Ni$_{0.035}$)$_2$As$_2$ single crystals. The difference between these values can be explained by the compressive strain exerted on the Ba(Fe$_{0.965}$Ni$_{0.035}$)$_2$As$_2$ films on CaF$_2$ substrate~\cite{Yoon}. This linear temperature dependence  testifies that the hidden transport property for the $x=0.035$ films indicates the non-Fermi-liquid behavior above and below the magnetic transition temperature. We also observe non-trivial temperature dependences for the two quantities that characterize the dc transport in the underdoped films, $\sigma _{dc}(T)\equiv\omega _{p}(T)^{2}/[4\pi\cdot\gamma _D(T)]$ and $\rho (T)\equiv 1/\sigma _{dc}(T)$ (see Figs. 5(c) and 5(d)). In Figs.~5(d) and 8(d) we display the total dc resistivity (triangles), which can be calculated using a relation $1/\rho _{D1+D2}=1/\rho _{D1}+1/\rho _{D2}$, and compare it with the results of Van der Pauw (VDP) measurements. There is a good agreement between optical and transport measurements, as evidenced also by the data for the remaining compositions (Figs.~6(d), 7(d), and 9(d)).

\begin{figure}[!ht]
\includegraphics[width=8.3cm]{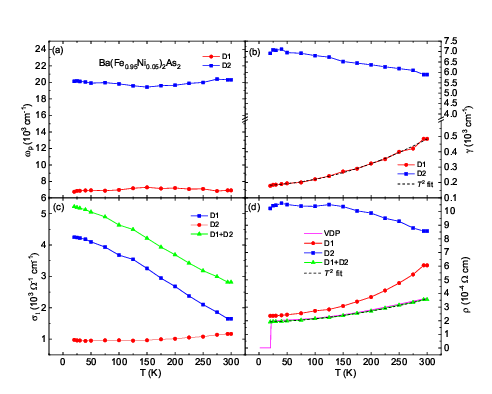}
\centering
\caption{(Color online) The fitting parameters $\omega _{Di,p}$ (a) and $\gamma _{Di}$ (b) of two Drude modes D$_i$ ($i=1,2$) of the Ba(Fe$_{0.95}$Ni$_{0.05}$)$_2$As$_2$ films. Also shown are the calculated dc conductivities $\sigma _{dc,i}(T)$ (c) and dc resistivities $\rho _{i}(T)$ (d) including the total conductivity and resistivity as functions of temperature. The solid purple line depicts the temperature dependence of resistivity obtained from Van der Pauw measurements. The dashed curves in the panels (b) and (d) are $T^2$ fits to the scattering rate for the narrow Drude component and to the total dc resistivity.}
%\label{Fig_6}
\end{figure}

In the case of the Ba(Fe$_{0.95}$Ni$_{0.05}$)$_2$As$_2$ and Ba(Fe$_{0.92}$Ni$_{0.08}$)$_2$As$_2$ films, the scattering rate data for the narrow Drude contribution demonstrate a quadratic temperature dependence, as can be clearly seen in Figs.~6(b) and 7(b) (9(b)), where the quadratic function is shown by the dashed line. The total dc resistivity estimated from the optical data also exhibits $T^2$ behavior, see Figs.~6(d) and 7(d) (9(d)). A quadratic temperature dependence is expected for electron-electron scattering, which should be dominant for correlated electron systems at low temperatures and subject to Landau's theory of a Fermi liquid~\cite{Abrikosov,Pines}.

\begin{figure}[ht]
\includegraphics[width=8.3cm]{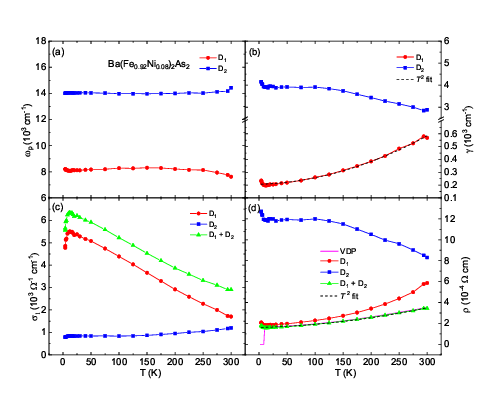}
\centering
\caption{(Color online) The fitting parameters $\omega _{Di,p}$ (a) and $\gamma _{Di}$ (b) of two Drude modes D$_i$ ($i=1,2$) of the Ba(Fe$_{0.92}$Ni$_{0.08}$)$_2$As$_2$ films obtained with the fixed Lorentz contributions. Also shown are the calculated dc conductivities $\sigma _{dc,i}(T)$ (c) and dc resistivities $\rho _{i}(T)$ (d) including the total conductivity and resistivity as functions of temperature. The solid purple line depicts the temperature dependence of resistivity obtained from Van der Pauw measurements. The dashed curves in the panels (b) and (d) are $T^2$ fits to the scattering rate for the narrow Drude component and total resistivity.}
%\label{Fig_7}
\end{figure}

\section{Resistivity in a multiband metal}

We take the temperature at which the derivative of the resistivity has a maximum as the critical temperature. We want to reproduce the temperature dependence of resistivity for a multiband superconductor. For this purpose, it is necessary to use the Allen theory ~\cite{Allen,Grimvall} generalized to a multiband system~\cite{LiFeAs,Aleshchenko2021,resumma,dolgov}, where we add up the contributions of all bands:
\begin{equation}
	\frac{1}{\rho(T)}=\frac{\varepsilon_0(\hbar\omega_{p})^{2}}{\hbar}\sum_{i=1}^N
						\frac{(\omega_{p,i}/\omega_{p})^2}{\Gamma_i+W_i(T)}.
\label{rho}
\end{equation}
In this equation, $\omega_{p,i}$ is the bare plasma frequency of the $i$-band, $\omega_{p}$ is the total bare plasma frequency, and
$N$ is the number of bands, while
\begin{equation}
	W_i(T)=4\pi k_BT\int_0^\infty d\Omega
			\left[\frac{\hbar\Omega/2k_BT}{\sinh\big(\hbar\Omega/2k_BT\big)}\right]^2
			\frac{\alpha_{tr,i}^2F_{tr}(\Omega)}{\Omega}.\\
\label{W1}
\end{equation}
In the last equations, we summarize the impurity contribution for each band: $\Gamma_{i}=\sum_{j=1}^N\Gamma_{ij}$, i.e., we sum the inter- and intra-band non-magnetic and magnetic impurity scattering rates. We do the same with the transport electron-boson spectral functions, where we summarize the inter- and intra-band contributions for each band.
\begin{equation}
\alpha_{tr,i}^{2}F_{tr}(\Omega)=\sum_{j=1}^N\alpha_{tr,ij}^{2}F_{tr}(\Omega),
\end{equation}
Here, $\alpha^2_{tr,ij}(\Omega)F_{tr}(\Omega)$ are the transport spectral functions related to the Eliashberg functions~\cite{Allen}. To highlight the value of the electron-boson coupling, we can write the spectral transport functions in this way: $\alpha_{tr,ij}^{2}F_{tr}(\Omega)=\lambda_{tr,ij}\alpha_{tr,ij}^{2}F^{n}_{tr}(\Omega)$,
where $\alpha_{tr,ij}^{2}F^{n}_{tr}(\Omega)$ is a normalized spectral function with electron boson coupling constant $\lambda_{n}=2\int_{0}^{+\infty}\frac{\alpha^{2}F^{n}(\Omega)}{\Omega}d\Omega$ equal to $1$.

To simplify our model, we assume, as we have already done for the SC state~\cite{Yuriisup}, that all shapes of spectral functions are equal. They differ just for a scaling factor, the coupling constant. When the coupling is mediated mainly by antiferromagnetic spin fluctuations (as in our case), this is a good approximation. Thus,
$\alpha_{tr,i}^{2}F_{tr}(\Omega)=\lambda_{tr,i}\alpha_{tr}^{2}F^{n}_{tr}(\Omega)$ and $\lambda_{tr,i}=\sum_{j=1}^N \lambda_{tr,ij}$.
The specific shape of the spectral function for the antiferromagnetic spin fluctuation in the normal state is standard \cite{Aleshchenko2021,resumma}.

The Fermi surface in the IBS presents a number of sheets variable from three to five, so a multi-band model is needed to explain SC and normal state properties. In this case, the number of free parameters become too large. To overcome this problem, we grouped all hole bands together and all electron bands together to investigate the electrical resistivity. Finally, we have a two-band model~\cite{dolgov} containing only two different kinds of carriers.

It's known that the electron-phonon coupling in all IBS is weak \cite{Lilia}, so the more plausible mechanism related to resistivity would be the antiferromagnetic spin fluctuations. From the Allen theory, it is known~\cite{Allen} that the SC and transport spectral functions are similar. The key difference manifests itself in the behavior for $\Omega\rightarrow 0$, where the transport function behaves like $\Omega^3$ instead of $\Omega$ as in the SC state. In the low-temperature superconductors the condition $\alpha^2_{tr}(\Omega)F_{tr}(\Omega)\propto\Omega^3$ takes place
in the range~\cite{Allen} $0 <\Omega< \Omega_{D}$, with $\Omega_{D}\approx\Omega_{0}/10$, while in the IBS the condition is the same but the range can be different and more large. The representative bosonic energy~\cite{Aleshchenko2021,resumma}, i.e., the peak of the spectral function is, of course, $\Omega_{0}$. In this way, the final definition of the transport spectral function is
$\alpha^2_{tr}(\Omega)F'_{tr}(\Omega)=b_{i}\Omega^{3}\vartheta(\Omega_{D}-\Omega)+c_{i}\alpha^2_{tr}(\Omega)F''_{tr}(\Omega)\vartheta (\Omega-\Omega_{D})$. The constants $b_{i}$ and $c_{i}$ have been fixed by requiring the continuity in $\Omega_{D}$ and the normalization. We chose for the $\alpha^2_{tr}(\Omega)F''_{tr}(\Omega)$ the functional form of the theoretical antiferromagnetic spin fluctuations function in the normal state~\cite{Popovich}:

\begin{equation}
\alpha_{tr}^{2}F''(\Omega)\propto\frac{\Omega_{0}\Omega}{\Omega^{2}+\Omega_{0}^{2}}\vartheta(\Omega_{c}-\Omega),
\label{a2F}
\end{equation}
where $\Omega_{c}$ is a cut-off energy (\mbox{$\Omega_{c}=1$ eV} in these calculations).

In principle, the transport spectral functions should depend on the temperature but we do not know precisely this dependence and therefore we used the shape for spin fluctuations in the normal state without the temperature dependence~\cite{dolgov}. As we said before, to reduce the number of free parameters, we use a two-band model, one consisting of holes and the other of electrons. Actually, within this model, we have a large number of parameters such as the electron-boson coupling constants $\lambda_{1,tr}$ and $\lambda_{2,tr}$, the impurities contents $\Gamma_1$ and $\Gamma_2$, the plasma energies $\omega_{p,1}$ and $\omega_{p,2}$, the representative energy $\Omega_{0}$, and the energy $\Omega_{D}$, connected with the behavior of spectral functions at low energies. In the case of Co-doped Ba122, ARPES and de Haas-van Alphen data indicate that the transport is drawn mainly by the electronic bands, while the hole bands are characterized by a weaker mobility~\cite{Maksimov}. This experimental fact means that the impurities are mostly concentrated in the hole band (indicated by the index 1), i.e. $\Gamma_1 \gg \Gamma_2$, while the transport coupling is much higher in the electron band (indicated by the index 2).

To use the simplest possible model, but still capture fundamental physics, we fix $\lambda_{1,tr}$ equal to zero. The consequence is that the contribution of one of the bands is independent of temperature, while the other changes the slope of the resistivity with the temperature.
Then from the acquired optical measurements (Fig.~6) we find $\omega_{p,1}=6717.5$~cm$^{-1}$ (837~meV) and $\omega_{p,2}=20101.6$~cm$^{-1}$ (2495~meV) for the case of films with nickel concentration $x=0.05$. The remaining free parameters are: $\Gamma_2$, $\Omega_{0}$, $\Omega_{D}$, and $\lambda_{2,tr}$ ($\Gamma_1$ is linked with the value of $\Gamma_1$ by $\rho(T=0)$) with the experimental conditions $\Gamma_1 \gg \Gamma_2$.

\begin{table*}[!ht]
\tiny
\centering
\begin{tabular}{|c|c|c|c|c|c|c|c|c|c|c|}
  \hline
%   after \\: \hline or \cline{col1-col2} \cline{col3-col4} ...
  $x$     & $\lambda_{2,tr}$ & $\lambda_{tr,tot}$ & $\Omega_{0}$ (meV)& $\Omega_{D}$ (meV) & $\Gamma_{2}$ (meV) & $\Gamma_{1}$ (meV) & $\omega_{p1}$ (meV) & $\omega_{p2}$ (meV) & $T_{c}$ (K) \\
  \hline
  0.035 & 0.23 &0.07 & 60 & 12 & 19.8 & 569.2  & 825  & 1896 & 20.27  \\
  0.05 & 0.37 &0.18 & 80 & 27 & 20.7 & 912.3  & 837  & 2495 & 20.36  \\
  0.080 & 0.29 &0.09 & 60 & 6  & 21.8 & 1852.7 & 1016 & 1737 & 9.27  \\
  \hline
\end{tabular}
\caption{Physical parameters obtained by fitting the experimental resistivity curves for the three different doping levels as a function of temperature. The uncertainty is in the last digit.}
\end{table*}

\begin{table*}[!ht]
\tiny
\centering
\begin{tabular}{|c|c|c|c|c|c|c|c|}
\hline
Compound & $T_c$, K & $\Delta _<$, & $\Delta _>$, & $2\Delta _</k_BT_c$ & $2\Delta _>/k_BT_c$ & Method & Reference\\
& & meV & meV & & & & \\
\hline
Ba(Fe$_{0.965}$Ni$_{0.035})_2$As$_2$ & 21.1 & 1.425 & 3.16 & 1.57 & 3.48 & Optical & ---\\
& & & & & & conductivity, & \\
& & & & & & this study & \\
\hline
Ba(Fe$_{0.93}$Co$_{0.07})_2$As$_2$ & 23 & 2.7--3.5 & 6.0--8.0 & 2.72--3.53 & 6.05--8.07 & Optical &  \\
& & & & & & conductivity & \cite{Heumen} \\
\hline
Ba(Fe$_{0.955}$Ni$_{0.045})_2$As$_2$ & 20 & 1.6--1.7 & 3.45--3.7 & 1.9--2.0 & 4.0--4.3 & Optical &  \\
& & & & & & conductivity & \cite{Aleshchenko2021} \\
\hline
Ba(Fe$_{0.95}$Ni$_{0.05})_2$As$_2$ & 21.6 & 1.75--1.95 & 3.4--3.84 & 1.88--2.08 & 3.66--4.13 & Optical & ---\\
& & & & & & conductivity, & \\
& & & & & & this study & \\
\hline
Ba(Fe$_{0.9}$Co$_{0.1})_2$As$_2$ & 20 & 1.85 & 3.95 & 2.15 & 4.6 & Optical & \cite{Maksimov}\\
& & & & & & conductivity & \\
\hline
Ba(Fe$_{0.95}$Pt$_{0.05})_2$As$_2$ & 23 & 1.95 & 3.6, & 1.97 & 3.63, & Optical & \cite{Xing 2018}\\
& & & 5.4 & & 5.44 & conductivity & \\
\hline
\end{tabular}
\caption{Some SC papameters for electron-doped Ba122 with the similar levels of doping.}
\end{table*}

\section{Discussion}

The resistivity as a function of temperature and the fitting curves for the films with three different Ni concentrations are displayed in Fig.~1. The fits agree quite well with the experimental data except in the case of $x=0.035$, where theory cannot reproduce the resistivity behavior near the critical temperature. In the underdoped case, the behavior of the type $exp(-\frac{\alpha}{k_{B}T})$ ($\alpha$ is a constant) is similar to that in a semiconductor, since at low doping our material is more similar to insulating parent compound. The same is observed for various HTSC cuprates~\cite{HTCS1}. The physical parameters obtained by fitting the experimental temperature dependences of resistivity for three different Ni contents are collected in Table~1. We can see that the values of $\Gamma_1$ and $\Gamma_2$ increase with the doping $x$, i.e., the disorder increases.

The inset to Fig. 1 show the spectral function of the antiferromagnetic spin fluctuations in the normal state normalized to unity. We can note that there is a big increase of the typical energy $\Omega_0$ of the electron-boson coupling from the SC to the normal state in agreement with inelastic neutron scattering data~\cite{Inosov} (in the SC state, $\Omega_{s0}=4.6 k_{B} T_{c}$~\cite{Talantsev2021}) and a significant decrease of the coupling constant $\lambda_{tr,tot}=\frac{N_{1}(0)\lambda_{tr,1}+N_{2}(0)\lambda_{tr,2}}{N_{1}(0)+N_{2}(0)}$ in the normal state ($N_{1}(0)$ is the normal density of states at the Fermi level in the hole-band, while $N_{2}(0)$ is the sum of the contribution of the two electronic bands) with respect to SC state $\lambda_{s,tot}=\frac{\sum_{ij}N_{i}(0)\lambda_{ij}}{\sum_{i}N_{i}(0)}\simeq 2$~\cite{Yuriisup}. This behavior is typical of other IBS~\cite{resumma,Charnu} and HTSC cuprates~\cite{Maxi}.

It can be clearly seen from Figs.~2(b) and 3(b) that the broad Drude term is not gapped at least in the frequency range of our measurements. For the case that the coherent transport in the normal state arises mainly from the electron pocket, the observed two energy scales for the SC gaps might be due to a  single anisotropic $s$-wave gap at the electron pocket (see the results for Co-doped Ba122 having very similar electronic structure~\cite{Tu 2010}). This picture is consistent with a $s_\pm $-wave pairing symmetry of the order parameter.

The characteristic ratios $2\Delta /k_{B}T_c$ obtained for the two gap values in the present study are 1.57 and 3.48 for the Ba(Fe$_{0.965}$Ni$_{0.035}$)$_2$As$_2$ films (we use $T_{c90}$ for comparison). These values are in reasonable agreement with $2\Delta /k_{B}T_c\approx 1.8$ and 4.3--5.9 found in~\cite{Kuzmicheva1} for the Ba(Fe$_{0.96}$Ni$_{0.04}$)$_2$As$_2$ single crystal using the  incoherent multiple Andreev reflection spectroscopy (IMARE). For the Ba(Fe$_{0.95}$Ni$_{0.05}$)$_2$As$_2$ films, the characteristic ratios are 1.88--2.08 and 3.66--4.13. These values correlate well with the SC gaps of single crystals of the same composition determined from specific heat measurements as well as directly by IMARE spectroscopy~\cite{Kuzmicheva2,Kuzmicheva3}.

A comparison of Figs.~5 and 7 with Figs.~8 and 9 in the Appendix allows us to conclude that regardless of the choice of approximations (fixed Lorentz contributions or fixed static scattering rate $\gamma _{D2}$), our Drude-Lorentz analysis of the optical conductivities for the Ba(Fe$_{0.965}$Ni$_{0.035}$)$_2$As$_2$ and Ba(Fe$_{0.92}$Ni$_{0.08}$)$_2$As$_2$ films in the normal state adequately describes the temperature dependences of $\gamma _{D1}$ and total resistivity. This lends further support to the reliability of the obtained results. The non-Fermi-liquid behavior above and below the magnetic transition temperature for the Ba(Fe$_{0.965}$Ni$_{0.035}$)$_2$As$_2$ films found by us was also observed earlier for the Ba(Fe$_{0.975}$Ni$_{0.025}$)$_2$As$_2$ single crystals~\cite{Lee}. It is also similar to that in the non-Fermi liquid phase of HTSC cuprates~\cite{Ito}. In addition, it should be noted that the magnetic phase transition in the underdoped films is closely related to the D$_1$ mode or electron carriers in the electron pocket. It seems to be consistent with earlier ARPES study of detwinned Ba(Fe$_{1-x}$Co$_x$)$_2$As$_2$ single crystals~\cite{Yi}. Our Ba(Fe$_{0.95}$Ni$_{0.05}$)$_2$As$_2$ and Ba(Fe$_{0.92}$Ni$_{0.08}$)$_2$As$_2$ films demonstrate $T^2$ behavior both for the static scattering rate $\gamma _{D2}$ and the total resistivity, i.e., Fermi-liquid behavior. Similar behavior was previously observed for the Ba(Fe$_{0.95}$Ni$_{0.05}$)$_2$As$_2$ films~\cite{Yoon} and Ba(Fe$_{0.955}$Ni$_{0.045}$)$_2$As$_2$ films~\cite{Aleshchenko2021} as well as for the Ba(Fe$_{0.95}$Ni$_{0.05}$)$_2$As$_2$ and BaFe$_{1.84}$Co$_{0.16}$As$_2$ single crystals~\cite{Lee,Wu1,Barisic}. In the contrary, for the hole-doped Ba$_{0.6}$K$_{0.4}$Fe$_2$As$_2$ single crystals a $T$-linear behavior was established~\cite{Dai}. This testifies that the electron- and hole-doped IBS possess different hidden D$_1$ transport properties.
{\sloppy

    }
Let us compare the magnitudes of SC gaps obtained in our studies for Ba(Fe$_{1-x}$Ni$_{x}$)$_2$As$_2$ with those extracted from the IR measurements for other electron-doped Ba122 systems such as Co- and Pt-doped Ba122 with the similar levels of doping. In Ba(Fe$_{1-x}$Ni$_{x}$)$_2$As$_2$ every Ni donates two 3$d$ electrons in contrast to Co doping that contributes only one 3$d$ electron. Thus one should use the compositions close to Ba(Fe$_{0.93}$Co$_{0.07})_2$As$_2$ and Ba(Fe$_{0.9}$Co$_{0.1})_2$As$_2$ for comparison (Table~2). In the case of Ba(Fe$_{0.95}$Pt$_{0.05})_2$As$_2$, three nodeless energy gaps were inferred from the optical conductivity data in the SC state~\cite{Xing 2018}. It can be concluded from Table~2 that the same quantity of similar dopant produces analogous effect on the $T_c$ and SC gap magnitudes in electron-doped Ba122 ferropnictides. This means that the mechanism of superconductivity in these materials is robust against the change of dopants with similar properties and not prone to fine tuning.{\sloppy

}
\section{Conclusions}

We have studied the optical and the hidden transport properties of the Ba(Fe$_{1-x}$Ni$_{x}$)$_2$As$_2$ films ($x=0.035$, 0.05, and 0.08) in the normal and SC states. A two-Drude model was found to describe successfully the optical properties of the films in the normal state. In the SC state, two gaps with $2\Delta _{0}^{(2)}/k_BT_c=1.57$ and $2\Delta _{0}^{(1)}/k_BT_c=3.48$ are formed for $x=0.035$ and $2\Delta _{0}^{(2)}/k_BT_c=1.88$--2.08 and $2\Delta _{0}^{(1)}/k_BT_c=3.66$--4.13 for $x=0.05$. Both SC gaps were found to be formed from the narrow Drude component, while the broad component remains ungapped. In the case of the Ba(Fe$_{0.92}$Ni$_{0.08}$)$_2$As$_2$ films, no gap is observed because the SC transition temperature $T_c$ is too low. The temperature dependences of the plasma frequencies, optical conductivities, scattering rates, and dc resisitivities of both Drude components in the normal state were examined. The hidden Fermi-liquid behavior found previously for optimally doped Ba(Fe$_{1-x}$Ni$_{x}$)$_2$As$_2$ films ($x\simeq 0.05$)~\cite{Yoon,Aleshchenko2021} and single crystals~\cite{Lee,Wu1} is confirmed for our Ba(Fe$_{0.95}$Ni$_{0.05}$)$_2$As$_2$ and Ba(Fe$_{0.92}$Ni$_{0.08}$)$_2$As$_2$ films. For the Ba(Fe$_{0.965}$Ni$_{0.035}$)$_2$As$_2$ films, the non-Fermi-liquid behavior was established. From temperature-dependent resistivity measurements, we found that the electron-boson coupling constant decreases strongly, while the representative boson energy increases strongly, as is also the case for other IBS materials~\cite{resumma,Golubov 2011} and HTSC cuprates~\cite{Maxi}. The similarity in behavior may stem from the fact that the mechanism responsible for superconductivity is, perhaps, similar for HTSC cuprates and IBS materials.

\section*{CRediT author statement}

\textbf{Yurii A. Aleshchenko}: Conceptualization, Writing - Original Draft. \textbf{Andrey V. Muratov}: Data Curation, Software, Formal analysis, Visualization. \textbf{Elena S. Zhukova}: Investigation. \textbf{Lenar S. Kadyrov}: Investigation. \textbf{Boris P.Gorshunov}: Writing - Original Draft, Writing - Review \& Editing, Funding acquisition. \textbf{Giovanni A. Ummarino}: Methodology, Investigation, Writing - Original Draft, Visualization. \textbf{Ilya A. Shipulin}: Resources, Investigation, Writing - Review \& Editing.

\section*{Declaration of competing interest}

The authors declare that they have no known competing financial interests or personal relationships that could have appeared to influence
the work reported in this paper.

\section*{Data availability}

Data will be made available on request.

\section*{Acknowledgements}

We thank Ruben H\"uhne for fruitful discussions and providing samples. G.A. Ummarino acknowledges partial support from the MEPhI. THz experiments were supported by the Ministry of Science and Higher Education of the Russian Federation (No. FSMG-2021-0005),  THz data processing and analysis was carried out with the support of the Ministry of Science and Higher Education of the Russian Federation (Grant No. 075-15-2024-632). The IR Fourier-transform and ellipsometric measurements were performed using research equipment of the Shred Facilities Center at LPI.

\section*{Appendix}

\begin{figure}[!ht]
\includegraphics[width=13.5cm]{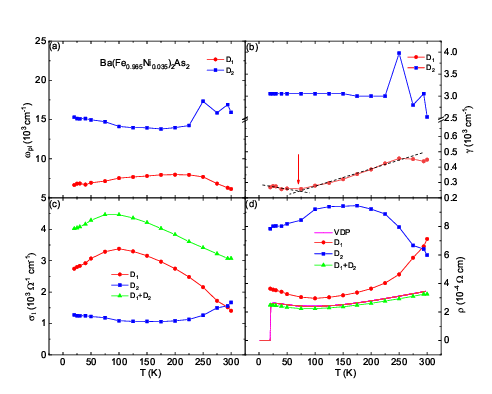}
\centering
\caption{(Color online) The fitting parameters $\omega _{Di,p}$ (a) and $\gamma _{Di}$ (b) of two Drude modes D$_i$ ($i=1,2$) of the Ba(Fe$_{0.965}$Ni$_{0.035}$)$_2$As$_2$ films obtained with the fixed static scattering rate $\gamma _{D2}=3000$~cm$^{-1}$. The dashed lines in the panel (b) are linear fits to the scattering rate for the narrow Drude component. The red arrow marks the magnetic transition. The calculated DC conductivities $\sigma _{dc,i}(T)$ (c) and DC resistivities $\rho _{i}(T)$ (d) including the total conductivity and resistivity as functions of temperature are shown. The solid purple line depicts the temperature dependence of resistivity obtained from Van der Pauw measurements.}
\label{Fig8}
\end{figure}

\begin{figure}[!ht]
\includegraphics[width=13.5cm]{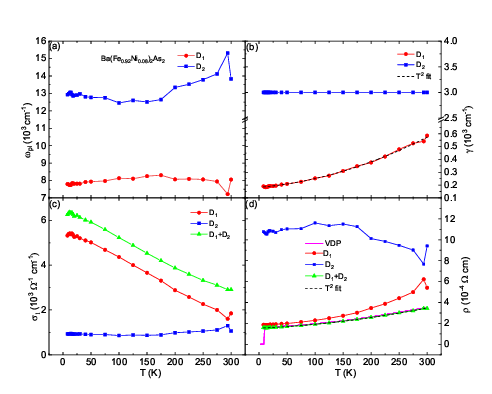}
\centering
\caption{(Color online) The fitting parameters $\omega _{Di,p}$ (a) and $\gamma _{Di}$ (b) of two Drude modes D$_i$ ($i=1,2$) of the Ba(Fe$_{0.92}$Ni$_{0.08}$)$_2$As$_2$ films obtained with the fixed static scattering rate $\gamma _{D2}=3000$~cm$^{-1}$. The calculated DC conductivities $\sigma _{dc,i}(T)$ (c) and DC resistivities $\rho _{i}(T)$ (d) including the total conductivity and resistivity as functions of temperature are shown. The solid purple line depicts the temperature dependence of resistivity obtained from Van der Pauw measurements. The dashed curves in the panels (b) and (d) are $T^2$ fits to the scattering rate for the narrow Drude component and to the total dc resistivity.}
%\label{Fig9}
\end{figure}

In the optical conductivity spectra of the Ba(Fe$_{0.965}$Ni$_{0.035}$)$_2$As$_2$ as well as Ba(Fe$_{0.92}$Ni$_{0.08}$)$_2$As$_2$ films, the D$_2$ Drude components become anomalously broad with decreasing temperature and overlap with the strong Lorentz contributions in the ranges of 3000--6000~cm$^{-1}$ and 13~500--16~500~cm$^{-1}$, thus preventing an unambiguous determination of the D$_2$ plasma frequency and damping in the course of the Drude-Lorentz analysis. This situation is illustrated in Figs.~2(b) and 4(b) of the main text. To overcome this problem, we fixed $\gamma _{D2}$ at 3000~cm$^{-1}$ in our calculations for the $x=0.035$ and $x=0.08$ films, which is comparable to that for the Ba(Fe$_{1-x}$Ni$_x$)$_2$As$_2$ single crystals with $x=0.025$ and $x=0.05$~\cite{Lee}, and films with $x=0.045$~\cite{Aleshchenko2021} and $x=0.05$~\cite{Yoon}. Moreover, $\gamma _{D2}$ has been proved to be temperature independent for the Ba(Fe$_{0.975}$Ni$_{0.025}$)$_2$As$_2$ and Ba(Fe$_{0.95}$Ni$_{0.05}$)$_2$As$_2$ single crystals~\cite{Lee} and Ba(Fe$_{0.95}$Ni$_{0.05}$)$_2$As$_2$ films~\cite{Yoon}, whereas it demonstrates only a weak temperature dependence below 200~cm$^{-1}$ in our previous work~\cite{Aleshchenko2021}. The results of the Drude-Lorentz modeling with fixed $\gamma _{D2}$ are shown in Figs.~8 and 9. Comparing Figs.~8 and 9 with corresponding Figs.~5 and 7 in the main manuscript one can conclude that the choice of the approximation (either fixed Lorentz contributions or fixed scattering rate of the broad Drude component $\gamma _{D2}=3000$~cm$^{-1}$) does not affect the results of the Drude-Lorentz analysis of the optical conductivities for the Ba(Fe$_{0.965}$Ni$_{0.035}$)$_2$As$_2$ and Ba(Fe$_{0.92}$Ni$_{0.08}$)$_2$As$_2$ films concerning the temperature dependences of $\gamma _{D1}$ and total resistivity in the normal state.

Figure~10 illustrates the behavior of the imaginary part of optical conductivity $\sigma _2$ for the films with $x=0.035$ (calculated with fixed $\gamma _{D2} = 3000$~cm$^{-1}$) and $x=0.05$. The divergence of the $\sigma _2$ inherent to the SC state is evident at the temperature of 4~K. Spectra of $\sigma _2$ for the Ba(Fe$_{0.92}$Ni$_{0.08}$)$_2$As$_2$ films are not presented since no upturn of reflectivity in the SC state was detected for this film. This may be due to the smaller SC gap for this Ni content and the limited frequency range of our measurements.

\begin{figure}[!ht]
\includegraphics[width=13.5cm]{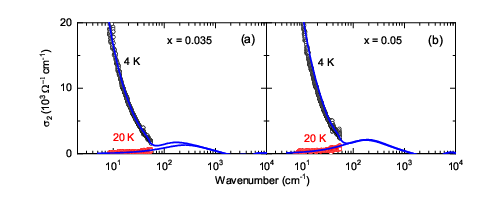}
\centering
\caption{(Color online) Spectra of the imaginary part of optical conductivity of the Ba(Fe$_{0.965}$Ni$_{0.035}$)$_2$As$_2$ (a) and Ba(Fe$_{0.95}$Ni$_{0.05}$)$_2$As$_2$ (b) films measured close to $T_c$ ($T=20$~K) and below it ($T=4$~K). Conductivity data below 55~cm$^{-1}$ (dots) were obtained with the THz time-domain measurements.}
%\label{Fig_10}
\end{figure}

\end{document}